\documentclass[aps,prl,superscriptaddress,twocolumn,showpacs,nofootinbib]{revtex4-1}
\usepackage{graphicx}
\usepackage{amsmath}
\usepackage{amssymb}
\usepackage{subfigure}
\usepackage{natbib}

\begin{document}

\title{Attaining sub-classical metrology in lossy systems with entangled coherent states}
\author{P.A. Knott}
	\email{phy5pak@leeds.ac.uk}
	\affiliation{School of Physics and Astronomy, University of Leeds, Leeds LS2 9JT, United Kingdom}
	\affiliation{NTT Basic Research Laboratories, NTT Corporation, 3-1 Morinosato-Wakamiya, Atsugi, Kanagawa 243-0198, Japan}
\author{W.J. Munro}
	\affiliation{NTT Basic Research Laboratories, NTT Corporation, 3-1 Morinosato-Wakamiya, Atsugi, Kanagawa 243-0198, Japan}
\author{J.A. Dunningham}
	\affiliation{Department of Physics and Astronomy, University of Sussex, Brighton BN1 9QH, United Kingdom}
\pacs{42.50.St,42.50.Dv,03.65.Ud,03.65.Ta,03.67}

\date{\today}

\begin{abstract}

Quantum mechanics allows entanglement enhanced measurements to be performed, but loss remains an obstacle in constructing realistic quantum metrology schemes. However, recent work has revealed that entangled coherent states (ECSs) have the potential to perform robust sub-classical measurements [J. Joo \textit{et. al.}, Phys. Rev. Lett. \textbf{107}, 83601 (2011)]. Up to now no read out scheme has been devised which exploits this robust nature of ECSs, but we present here an experimentally accessible method of achieving precision close to the theoretical bound, even with loss. We show substantial improvements over unentangled `classical' states and highly-entangled NOON states for a wide range of loss values, elevating quantum metrology to a realizable technology in the near future.

\end{abstract}
\maketitle


Quantum metrology aims to harness the power of quantum mechanics to make ultra-precise measurements \citep{caves1982quantum}. This has many important applications \citep{giovannetti2004quantum} including gravitational wave detection \citep{aasi2013enhanced,schnabel2010quantum}, quantum lithography \citep{d2004two,boto2000quantum}, and biological sensing \citep{taylor2013biological,taylor2013sub,nagata2007beating,anisimov2010quantum}. A crucial advantage of quantum metrology is in providing comparable precision with a significantly lower particle flux, an important requirement for many of these technologies such as in biological sensing \citep{crespi2012measuring}, where disturbing the system can damage the sample, or in gravitational wave detection, where the lasers in the interferometer interact with the mirrors enough to degrade the measurement \citep{purdy2013observation,harry2010advanced,goda2008quantum}. Quantum metrology is also a stepping stone towards more advanced quantum technologies as state preparation, manipulation and measurement are common requirements of technological applications of quantum theory \citep{dowling2003quantum,o2009photonic,nielsen2010quantum}. Furthermore, measurements are fundamental in physics and future success depends in part on the effectiveness of the measuring devices available.

It is known that an interferometer that utilizes a stream of independent particles is capable of measurement precision at the shot noise limit (SNL) $ 1/{\sqrt{n}}$ \citep{gkortsilas2012measuring} where $n$ is the total number of particles used in the probe state. However, by making use of quantum mechanical properties this can be improved to the ``Heisenberg limit" $1/{n}$ \citep{sanders1995optimal,dunningham2006using,xiang2010entanglement,boto2000quantum}. The problem with such an approach is that quantum states are notoriously fragile to particle losses \citep{rubin2007loss}, which typically collapse a state and destroy the phase information. A number of clever schemes have been devised with some robustness to loss which still capture sub-classical precision such as the NOON `chopping' strategy \citep{dorner2009optimal}, unbalanced NOON states \citep{demkowicz2009quantum} and BAT states \citep{gerrits2010generation}. While these states achieve sub-classical precision with a small amount of loss, for realistic losses likely to be experienced in an experiment they soon lose their advantage and are beaten by unentangled measurement schemes.

A class of states that show the potential for a great improvement over these alternatives are the entangled coherent states (ECSs) \citep{gerry1997generation,munro2001weak,gerry2009maximally,gerry2010heisenberg,sanders2012review,joo2012quantum}. Indeed Joo \textit{et. al.} \citep{joo2011quantum} used the quantum Fisher information (QFI) to show that ECSs can beat unentangled states and NOON states for most loss rates -- including the higher loss rates inaccessible by other schemes. Nevertheless a big problem that has stalled this avenue for advancement is the lack of any read out that can use an ECS to measure a phase to a high precision when loss is present. We present here a scheme that overcomes the former difficulties to attain precision close to the ultimate limit given by the QFI in a lossy system. Furthermore all the steps of this scheme are feasible with current technologies or technologies in the near future, demonstrating that sub-classical measurements robust to loss are realistically achievable.

The QFI for a general state $\rho$ is given by \citep{braunstein1994statistical,boixo2009quantum,luo2004wigner}:
\begin{eqnarray}
F_Q = \text{Tr} \left( \rho A^2 \right)
\end{eqnarray}
where $A$ is found from solving the symmetric logarithmic derivative $\partial\rho/\partial\phi = 1/2 \left[ A \rho + \rho A \right]$. The precision in the phase measurement (more specifically the lower bound on the standard deviation) is given by the quantum Cram\'er-Rao bound \citep{braunstein1994statistical}:
\begin{eqnarray}
\delta \phi \ge \frac{1}{\sqrt{\mu F_Q}},
\end{eqnarray}
where $\mu$ is the number of times that the measurement is independently repeated.
This gives the best possible precision with which a state can measure a phase. For NOON states and unentangled states the quantum Cram\'er-Rao bound gives us the Heisenberg and shot noise limits respectively \citep{durkin2007local,helstrom1976quantum}. Joo \textit{et. al.} \citep{joo2011quantum} used the QFI to numerically show that with and without loss the ECS can achieve better precision than unentangled, NOON, and some other candidate states. Zhang \textit{et. al.} \citep{zhang2013quantum} added to this by deriving an expression for the QFI with loss for arbitrary amplitude $\alpha$, and confirmed the potential of ECSs for robust quantum metrology. They formulated the QFI for ECSs as being comprised of two parts so that $F_Q=F_H+F_{cl}$, where $F_H$ represents the part of the state that allows Heisenberg limited precision, while $F_{cl}$ observes classical SNL precision. This reveals the power of ECSs: unlike other quantum states the ECSs can retain at least SNL precision with high loss, a feature that should be, but has not yet been, exploited.

\begin{figure}
\centering
\includegraphics[scale=1.32]{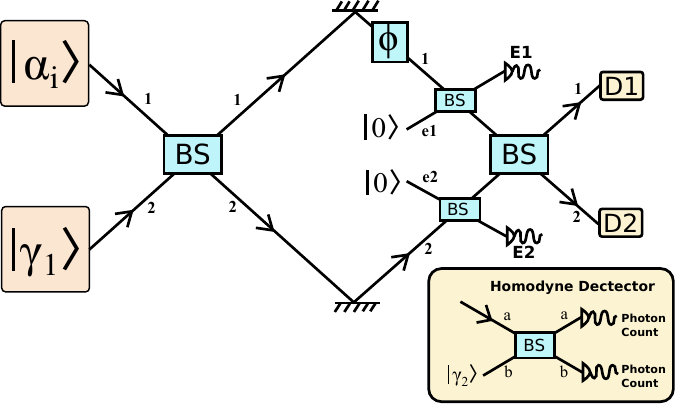}
\caption{ (Color online) The general scheme for measuring a phase using an entangled coherent state. Loss is modeled by the addition of a beam splitter in each arm (with a vacuum port). The final detectors $D1$ and $D2$ can be chosen to maximize precision as discussed in the text. The inset shows the homodyne measurement with reference state $|\gamma_2>$ that we will use for $D1$ and $D2$ when there is loss.}
\label{fig:ECS_longarm_general_concise3}
\end{figure}


\emph{Simple scheme without loss -} Fig.~\ref{fig:ECS_longarm_general_concise3} illustrates the interferometer we employ. The first task is to create the ECS \citep{munro2000entangled}, and for this we will use the method proposed by Gerry \textit{et. al.} \citep{gerry2009maximally}. We take for the lower input an even cat state \citep{gerry1993generation,gerry1993non,tilma2010entanglement,ralph2003quantum}, which contains only even numbers of photons: $|\gamma_1\rangle=\mathcal{N}_1(|\alpha_0/\sqrt{2}\rangle+|-\alpha_0/\sqrt{2}\rangle$ where $\mathcal{N}_1=1/\sqrt{2+2e^{-|\alpha_0|^2}}$ and for the upper input we use a coherent state $|\alpha_i\rangle=|\alpha_0/\sqrt{2}\rangle$. After the first 50:50 beam splitter in Fig.~\ref{fig:ECS_longarm_general_concise3} we then have the ECS:
\begin{align}
|\psi_1 \rangle_{1,2} = \mathcal{N}_1 \left( |\alpha_0,0 \rangle + |0,\alpha_0 \rangle \right).
\end{align}

Other methods of creating an ECS have been proposed but these schemes require nonlinear interferometers \citep{gerry2009maximally,sanders1992entangled} which are tough to make experimentally as a strong source of Kerr nonlinearity is required. Producing the cat state $|\gamma_1\rangle$, which we use to create our ECS, is likely to be easier in comparison \citep{ourjoumtsev2007generation,takahashi2008generation}. For example, in \citep{brune1996observing} a superposition is made of a Rydberg atom in a cavity: $|g \rangle + |e \rangle$. A coherent state is then introduced and the Jaynes-Cummings Hamiltonian \citep{jaynes1963comparison} is applied to give $|\alpha \rangle (|g \rangle + |e \rangle) \rightarrow |\alpha \rangle |g \rangle + |\alpha e^{i\phi} \rangle |e \rangle$. The Rydberg atom is then transformed and measured, and if we take $\phi=\pi$ we are left with the even cat state. These methods for creating cat states also need an effective non linearity, but the important advantage here over nonlinear interferometers is that the cat state is created offline whereas it is necessary to implement the nonlinearity of the interferometer within the scheme itself. In principle we could have a device that waits for a cat state to be successfully created, and then inputs it into the interferometer to be used for the phase estimation.

We first consider a parity measurement scheme, which performs well without loss \citep{gerry2010heisenberg,joo2011quantum,gerry2000heisenberg}. After creating the ECS we apply a linear phase shift $\phi$ to mode 1, giving:
\begin{align}
\label{equ:ECS_phased}
|\psi_1 \rangle \xrightarrow{\rm U=e^{i\phi(a_1^{\dagger}a_1^{})}} |\psi_2 \rangle = \mathcal{N}_1 \left( |\alpha_0 e^{i\phi},0 \rangle + |0,\alpha_0 \rangle \right).
\end{align}
Ignoring loss for the moment, the next step is to send this state through a second beam splitter which gives us the state:

\begin{align}
\mathcal{N}_1 e^{-\frac{|\alpha_0|^2}{2}} \sum_{m,n=0}^{\infty} \frac{\alpha_0^{m+n} \left( e^{i(m+n)\phi} + (-1)^n \right)}{\sqrt{2^{m+n}m!n!}}|m,n\rangle
\end{align}
where $m$ and $n$ are the number of photons incident on detectors $D1$ and $D2$, respectively, which we take here to be photon number resolving detectors. The probability of detecting $(m,n)$ photons at $(D1,D2)$ is then given by:
\begin{eqnarray}   
\label{eq:Prob_phi_given}
P(m,n) \propto 
\begin{cases}
\cos^2{(m+n)\phi \over 2}  & \text{for $n$ even} \\ \\
\sin^2{(m+n)\phi \over 2}  & \text{for $n$ odd}.
\end{cases}
\end{eqnarray}

We therefore must know if the output at detector $D2$ is even or odd in order to determine the phase (this will become important when we introduce loss). It can be shown that this scheme, with no loss, allows us to beat the best possible precision obtainable using NOON states of comparable sizes and also unentangled states\footnote{We take equivalently sized \uppercase{NOON} and \uppercase{ECS} states so that $\uppercase{N} =2\mathcal{N}_1^2 |\alpha_0|^2$.}. For small $\alpha_0$ we do not saturate the QFI, but we significantly improve upon the best possible measurement using a NOON state \citep{joo2011quantum}. For large $\alpha_0$ this scheme comes very close to saturating the QFI, but it is shown in \citep{joo2011quantum} that that in this region ECSs show only a small advantage over NOON states.


\emph{Introducing loss -} We model loss by the addition of beam splitters after the phase shift \citep{gkortsilas2012measuring,joo2011quantum,demkowicz2009quantum} as shown in Fig.~\ref{fig:ECS_longarm_general_concise3}, which have probability of transmission $\eta$, and therefore the fraction of the population lost is $\mu=1-\eta$. After these beam splitters we have the state:
\begin{align*}
\label{}
|\psi_1 \rangle_{e1,1,2,e2} &= \\ &\mathcal{N}_1 \left[ |\alpha_{0\mu} e^{i\phi},\alpha_{0\eta} e^{i\phi},0,0\rangle +  |0,0,\alpha_{0\eta},\alpha_{0\mu} \rangle \right].
\end{align*}
where $\alpha_{0\eta}=\alpha_0\sqrt{\eta}$ and $\alpha_{0\mu}=\alpha_0\sqrt{\mu}$. Tracing over the loss modes $e1$ and $e2$ we get:

\begin{align*}
\label{}
\rho = &\mathcal{N}_1^2 ( |\alpha_{0\eta} e^{i\phi},0 \rangle \langle \alpha_{0\eta} e^{i\phi},0 | + |0,\alpha_{0\eta} \rangle \langle 0,\alpha_{0\eta} | \\ &+ e^{-|\alpha_{0\mu}|^2} \left[ |\alpha_{0\eta} e^{i\phi},0 \rangle \langle 0,\alpha_{0\eta} | + |0,\alpha_{0\eta} \rangle \langle \alpha_{0\eta} e^{i\phi},0 | \right] ).
\end{align*}

\begin{figure}
\centering
\begin{subfigure}[]{}
\includegraphics[scale=0.437]{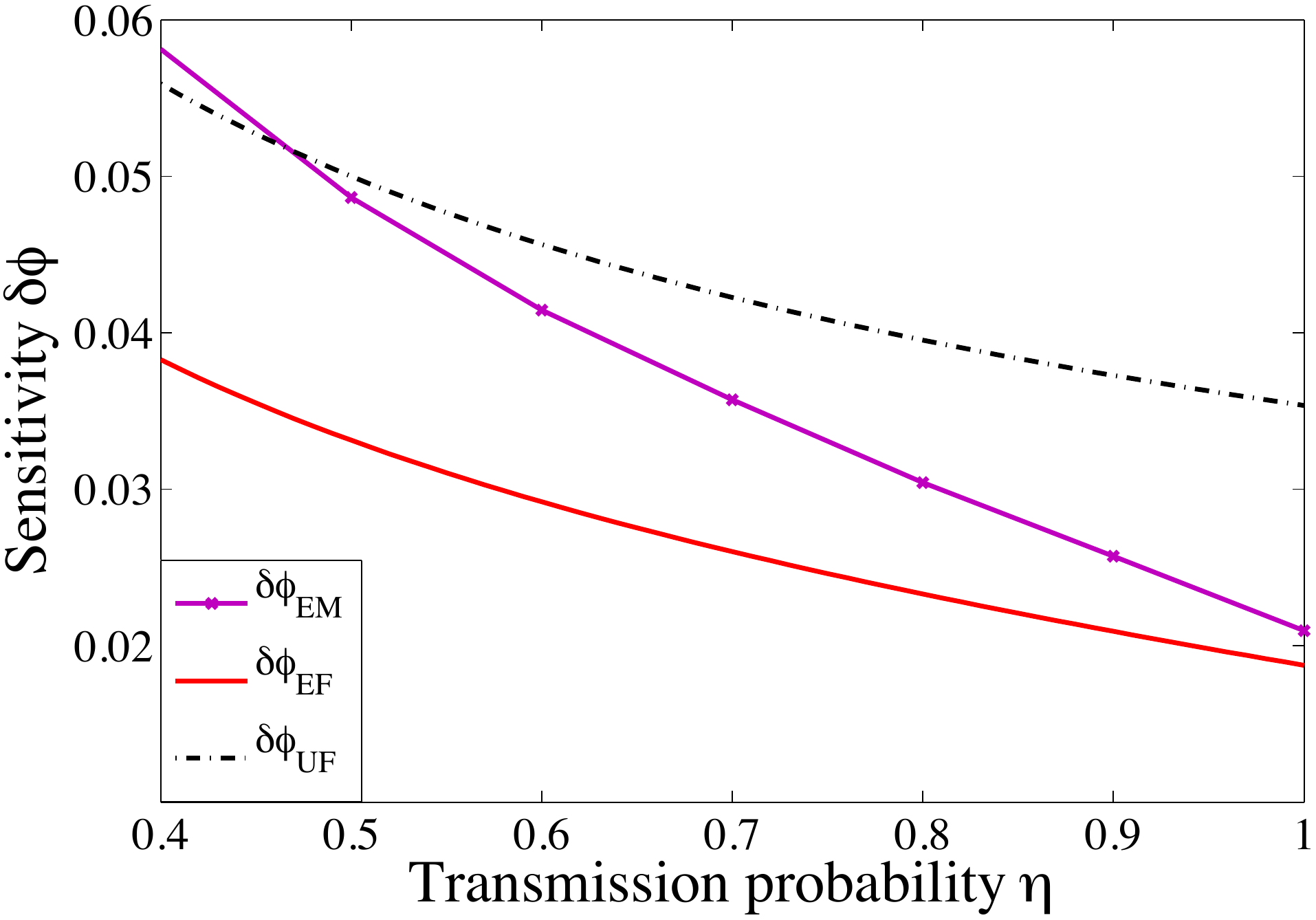}
\end{subfigure}
\centering
\begin{subfigure}[]{}
\includegraphics[scale=0.475]{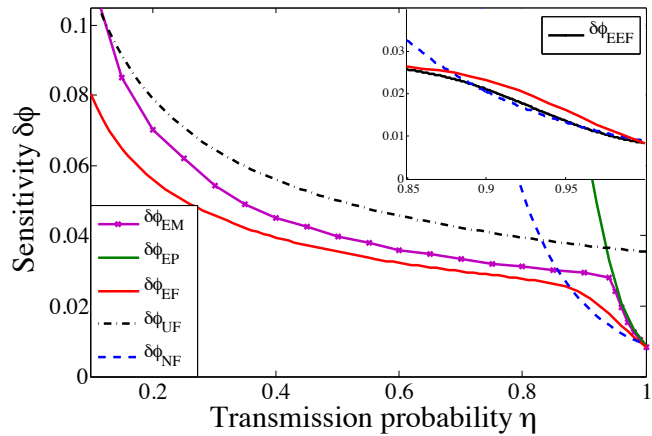}
\end{subfigure}
\caption{ (Color online) The measurable phase precision for ECSs with amplitudes (a) $\alpha_0=1.1307$ (which has an average photon number of 1) and (b) $\alpha_0=4$ using our measurement scheme are shown by the purple crossed lines $\delta\phi_{EM}$. The red solid, blue dashed and black dotted-dashed lines give the QFI of the ECS $\delta\phi_{EF}$, the NOON state $\delta\phi_{NF}$, and unentangled states $\delta\phi_{UF}$, respectively, all of equivalent size: $\uppercase{N} =2\mathcal{N}_1^2 |\alpha_0|^2$ (therefore in (a) the NOON and unentangled states are the same). In (b) the dark green solid line shows the simple parity measurement of the ECS $\delta\phi_{EP}$. For both small and large $\alpha_0$ our scheme provides the best phase precision for the majority of loss rates, and for larger $\alpha_0$ we come close to saturating the QFI. The black solid line in the inset in (b) shows the QFI of the even ECS, demonstrating how we can obtain a higher precision than the NOON states for most loss rates simply by modifying our input state. The other lines in the inset are the same as in the main figure.}
\label{fig:al4_best_opt}
\end{figure}

This mixed state points to a potential problem when using an ECS for metrology. The exponential suppression of the off diagonal coherence of the ECS, combined with the diminished $\alpha_{0\eta}$, leads to a rapid drop in phase precision for small loss. The result of this is that with the simple parity measurement scheme discussed above and used in \citep{joo2011quantum} the ECSs lose phase precision with loss significantly faster than NOON states. This is shown in Fig.~\ref{fig:al4_best_opt}(b) for $\alpha_0=4$, where the blue dashed line is for NOON states $\delta\phi_{NF}$, and the dark green solid line is for the ECS with the parity measurement $\delta\phi_{EP}$.

Nonetheless $\rho $ can be written in another form which reveals the great advantage of ECSs:

\begin{align}
\label{}
\rho = c_1 | \psi_+ \rangle \langle \psi_+ | + c_2 | \psi_- \rangle \langle \psi_- |
\label{rho2}
\end{align}
where $c_{12}={1 \over 2}(1 \pm e^{-|\alpha_{0\mu}|^2})$ and:
\begin{eqnarray}
|\psi_{\pm} \rangle = \mathcal{N}_1 \left[ |\alpha_{0\eta} e^{i\phi},0\rangle \pm |0,\alpha_{0\eta} \rangle \right].
\end{eqnarray}

The resulting state is a mixture of two pure ECSs which both contain phase information. If this were to be compared with the mixed state for a NOON state with loss, then one would find that the loss component of the NOON state contains no phase information. Despite this, the simple parity measurement scheme discussed above cannot determine $\phi$ when there is loss and this just contributes noise to the signal from the no-loss state.


\emph{Robust scheme with loss -} We will now present a scheme that can be used to recover the lost phase information that has so far eluded measurement. The key is to use extra reference coherent states above and below the main interferometer which can be used to perform homodyne measurements and recover the phase information. The measurement scheme is shown in the inset of Fig.~\ref{fig:ECS_longarm_general_concise3}. For measurement $D1$ we simply take $|\gamma_2\rangle_{1b}=|\alpha_1\rangle$. But due to the probabilities in Eq.~(\ref{eq:Prob_phi_given}) we must know whether even or odd numbers of particles are output at $D2$. Thus if we mix this state with $|\gamma_2\rangle_{2b}=|\alpha_1\rangle$ then, as we don't know the number of photons in a coherent state, we no longer know if the state had even or odd numbers. In this case the Heisenberg limited phase information provided by the entangled state washes out and we no longer get quantum enhancement (we can still measure the phase but at the shot noise limit at best). We therefore take $|\gamma_2\rangle_{2b}=\mathcal{N}_2(|\alpha_1\rangle+|-\alpha_1\rangle)$ where $\mathcal{N}_2=1/\sqrt{2(1+e^{-2|\alpha_1|^2})}$, which always contains an even number of photons, and therefore allows us to retain quantum enhancement.

The state directly after the phase shift, when we include the reference states used in the final detectors, is:
\begin{align*}
|\Psi \rangle_{1b,1,2,2b} = \mathcal{N}_1 \left( |\alpha_1,\alpha_0 e^{i\phi},0,\gamma_2 \rangle + |\alpha_1,0,\alpha_0,\gamma_2 \rangle \right).
\end{align*}
After loss we have the state:
\begin{align}
\label{}
\rho &= |\Phi_{1\eta} \rangle \langle \Phi_{1\eta} | + |\Phi_{2\eta} \rangle \langle \Phi_{2\eta} | \\ &+ e^{-|\alpha_{0\mu}|^{2}} \left( |\Phi_{1\eta} \rangle \langle \Phi_{2\eta} | + |\Phi_{2\eta} \rangle \langle \Phi_{1\eta} | \right),
\end{align}
where $|\Phi_{1\eta}\rangle=\mathcal{N}_1 |\alpha_{1} , \alpha_{0\eta}e^{i\phi}, 0,  \gamma_2 \rangle$ and $|\Phi_{2\eta}\rangle=\mathcal{N}_1 |\alpha_1 , 0,  \alpha_{0\eta},  \gamma_2 \rangle$. We can then send $\rho_2$ through the remainder of the interferometer, giving the probabilities at the outputs as:
\begin{align*}
\label{}
P(\#) &= \langle \# |\overline{\Phi_{1\eta}} \rangle \langle \overline{\Phi_{1\eta}} | \# \rangle + \langle \# |\overline{\Phi_{2\eta}} \rangle \langle \overline{\Phi_{2\eta}} | \# \rangle \\ &+ e^{-|\alpha_{0\mu}|^{2}} \left[ \langle \# |\overline{\Phi_{1\eta}} \rangle \langle \overline{\Phi_{2\eta}} | \# \rangle + \langle \# |\overline{\Phi_{2\eta}} \rangle \langle \overline{\Phi_{1\eta}} | \# \rangle \right],
\end{align*}
where $|\#\rangle = |k,l,m,n\rangle_{1b,1a,2a,2b}$, the state with $k$ particles in the first number resolving detector, $l$ in the second and so on. The barred states $|\overline{\Phi_{1\eta}}\rangle$ and $|\overline{\Phi_{2\eta}}\rangle$ can be found by sending $|{\Phi_{1\eta}}\rangle$ and $|{\Phi_{2\eta}}\rangle$ through the remainder of the interferometer.

We must then optimize over $\alpha_1$ to find the best precision these scheme can achieve: for different loss rates it is advantageous to use different sized reference states. The precision with which this scheme can measure the phase also depends on the (approximate) phase being measured $\phi$, as is true for most schemes, but it is relatively insensitive to the phase over a significant range. Nonetheless this should not pose much of a problem as we can just put a variable phase shift in mode 2, which allows us to vary the phase difference so that effectively $\phi$ can be whatever we choose. After optimizing over $\alpha_1$ and $\phi$ we then obtain the results in Fig.~\ref{fig:al4_best_opt}(a) for $\alpha_0=1.1307$ (which has an average photon number of 1). It can be seen that our state now out-performs the NOON and unentangled states for all values of loss up to $\eta=0.47$. The significant precision enhancement for small $\alpha_0$ is evident here, as well as the robustness to loss. Fig.~\ref{fig:al4_best_opt}(b) then shows the results for the larger amplitude ECS of $\alpha_0=4$. We can see that for larger $\alpha_0$ our scheme beats the competitors for most $\eta$ values. We note here that our scheme does not beat the SNL in scaling, as this is impossible when any loss is present \citep{escher2011general,demkowicz2012elusive}. What we do show is that when modest particle numbers are used our scheme can provide a more precise measurement than what is possible using uncorrelated states.

Fig.~\ref{fig:al4_best_opt}(b) also illustrates the agreement between our precision measurement and the Fisher information given by Zhang \textit{et. al.} \citep{zhang2013quantum}, shown as the solid red line $\delta\phi_{EF}$. We can see that our scheme, in agreement with the Fisher information, loses out to the NOON states initially, but before long our scheme exploits the presence of the phase in the loss terms and shows great improvement over the NOON and unentangled states for most loss values. Furthermore, for larger amplitude $\alpha_0$ we come close to the ultimate precision for the ECS given by the Cram\'er-Rao bound. The vast improvement of our scheme over the parity measurement is clear in Fig.~\ref{fig:al4_best_opt}(b), revealing the precision gained in including the extra reference states in the measurement.

We now show how our scheme can be improved further to overcome the rapid initial loss of coherence which results in our state losing out to the NOON state in the small loss regime. If we change our upper input state $|\alpha_i \rangle$ to another cat state $|\gamma_1\rangle=\mathcal{N}_1(|\alpha_0/\sqrt{2}\rangle+|-\alpha_0/\sqrt{2}\rangle$ then after the first beam splitter we will have an ECS that only contains even numbers of photons. The QFI for this even ECS is shown in the inset of Fig.~\ref{fig:al4_best_opt}(b): we now only marginally lose to the NOON state, in a very small region. This significant improvement in precision for low loss is due to a reduced suppression of the off diagonal coherence. We can now tailor our input states for different loss values to produce a scheme that achieves higher precision than NOON states and unentangled states for the vast majority of loss rates, including the experimentally relevant rates which can be up to a few times $10\%$ \citep{demkowicz2013fundamental}.

\emph{Conclusion -} Up to now it has not been at all clear how the full potential of ECSs as robust states for quantum metrology, as demonstrated by their QFI, can be exploited. Previous measurement schemes were unable to access the full phase information stored in the ECS after loss, and the suppression of the off diagonal coherence had the effect of making ECSs even worse than NOON states. However, we have presented here a more advanced measurement scheme that not only recovers the phase information with loss, but also comes close to saturating the QFI. Moreover we have shown that the input can be tailored so that we can always achieve higher precision than the NOON state. This allows us to achieve sub-classical precision measurements that outperform the alternative states for the majority of loss rates, including the rates thought to be realistic in an experiment. Furthermore, our scheme uses quantum resources that have already been created in the lab, bringing entanglement enhanced measurements in lossy systems within reach of current technology.

\begin{acknowledgments}
This work was partly supported by DSTL (contract number DSTLX1000063869).
\end{acknowledgments}

\bibliographystyle{apsrev}
\bibliography{MyLibrary}

\end{document}